\documentclass[journal]{IEEEtran}
\usepackage{xr}
\usepackage{amsthm}
\usepackage[mathscr]{eucal}
\usepackage{amsfonts}
\usepackage{amsmath}
\usepackage{amssymb}
\usepackage{graphicx}
\usepackage{xcolor}
\usepackage{comment}
\usepackage{pstricks}
\usepackage{fullpage}
\usepackage{hyperref}
\usepackage{booktabs}
\usepackage{colortbl}
\usepackage{enumitem}
\usepackage{setspace}
\usepackage{tabularx}
\usepackage{color}
\usepackage{tikz}

\usepackage{pdflscape}
\usepackage{floatrow}
\usepackage{subfig} 

\newcolumntype{N}{>{\centering\arraybackslash}m{1in}}

\newcolumntype{R}{>{\raggedright\arraybackslash}m{0.65in}}
\newcolumntype{Q}{>{\centering\arraybackslash}m{0.82in}}
\newcolumntype{P}{>{\raggedleft\arraybackslash}m{0.1125in}}
\newcolumntype{S}{>{\centering\arraybackslash}m{1.25in}}
\newcolumntype{T}{>{\centering\arraybackslash}m{0.5in}}
\newcolumntype{V}{>{\raggedleft\arraybackslash}m{0.9in}}

\newrgbcolor{gray70}{0.7019608 0.7019608 0.7019608}
\newrgbcolor{orangeR}{1 0.6470588 0}
\newrgbcolor{red4}{0.545098 0 0}
\newrgbcolor{olivedrab}{0.4196078 0.5568627 0.1372549}
\newrgbcolor{royalblue}{0.2549020 0.4117647 0.8823529}

\newcommand{\wh}[1]{\widehat{#1}}

\begin{document}
%
\title{On the potential of BFAST for monitoring burned areas using multi-temporal 
Landsat-7 images}
\author{\IEEEauthorblockN{Inder~Tecuapetla-G\'omez,
Gabriela~Villamil-Cortez and Mar\'ia Isabel~Cruz-L\'opez}
\thanks{
I.~Tecuapetla-G\'omez G.~Villamil-Cortez and M.~I.~Cruz-L\'opez,
Subdirecci\'on de Percepci\'on Remota,
National Commission for the Knowledge and Use of Biodiversity
(CONABIO), Liga Perif\'erico-Insurgentes Sur 4903,
Parques del Pedregal, Tlalpan 14010, Ciudad de M\'exico.

I.~Tecuapetla-G\'omez,
Direcci\'on de C\'atedras,
National Council of Science and Technology (CONACyT),
Av.~Insurgentes Sur 1582, Cr\'edito Constructor, 3940, Benito Ju\'arez,
Ciudad de M\'exico (e-mail: itecuapetla@conabio.gob.mx).
}}

\maketitle

\begin{abstract}
In this letter, we propose a semi-automatic approach 
to map burned areas and assess burn severity that does
not require prior knowledge of the fire date.
First, we apply BFAST to NDVI time series and estimate statistically abrupt 
changes in NDVI trends. 
These estimated changes are then used as plausible fire dates 
to calculate dNBR following a typical pre-post fire assessment.
In addition to its statistical guarantees, 
this method depends only on a tuning parameter
(the bandwidth of the test statistic for changes).
This method was applied to Landsat-7 images taken over
La Primavera Flora and Fauna Protection Area, in Jalisco, Mexico, from 2003 to 2016. 
We evaluated BFAST's ability to estimate vegetation changes
based on time series with significant observation 
gaps.
We discussed burn severity maps associated with
massive wildfires (2005 and 2012) and another with
smaller dimensions (2008) that might have been 
excluded from official records.
We validated our 2012 burned area map against a high resolution
burned area map obtained from RapidEye images;
in zones with moderate data quality, the overall accuracy of our map is $92\%$.
\end{abstract}


%
\IEEEpeerreviewmaketitle

\section{Introduction}~\label{sec.intro}

\IEEEPARstart{C}{orrectly} assessing wildfires is an effective
means of contributing to the protection of forests
and halting biodiversity loss.
An indirect and economical way 
to achieve the latter is to gather
satellite-derived products 
and apply to them a sounding method
for mapping burned areas and, subsequently, 
quantify possible severity or regrowth levels.

When identifying burned areas, three basic aspects should be considered:
1) presence of fuel (in this case, vegetation);
2) abrupt changes in certain spectral indices; and 
3) persistence of the abrupt change over time, cf.~\cite{chuvieco2008global}. 
Following these principles, several strategies and spectral
indices have been used to map burned areas 
based on Landsat imagery (see \cite{stroppiana2011positive},
\cite{Hawbaker.etal.2017}
\cite{zhao2015use}, 
\cite{roy2019landsat} and \cite{campagnolo2019patch} for some related work).
In some regions, like the one considered in this letter, 
Landsat-7 data provides the
best spatial (30m) and temporal resolution for vegetation monitoring in general.
As a result of the SLC failure of May 2003, however, any analysis using
Landsat-7 
must factor in an average loss of $22\%$ of the scene's data.

In this letter, we propose the application of
\cite{verbesselt2010detecting}'s BFAST (Breaks For Additive Seasonal Trend)
to NDVI and NBR time series of Landsat-7
imagery for burned area mapping, burn severity assessment and long-term monitoring.
We also evaluate BFAST's ability to detect
vegetation changes in time series with 
significant observation gaps.
The proposed methodology is applied to data taken over 
La Primavera Flora and Fauna Protection Area
in Jalisco, Mexico between 2003 and 2016.
Covering an area of
30,500ha, La Primavera is considered a lung for the
metropolitan area of Guadalajara, the third largest city in
Mexico.
A copy of the code employed for our analysis can be found
at the GitHub repository 
\href{https://github.com/inder-tg/burnSeverity}{https://github.com/inder-tg/burnSeverity}.

\section{Methodology}~\label{sec.methods}
BFAST statistically estimates
abrupt changes in the trend
structure of seasonally-driven time series.
More precisely, if
$y_t$ denotes the value of an NDVI time series at time $t$,
BFAST assumes an 
additive representation for $y_t$:
\begin{equation}~\label{eq.BFAST.model}
	y_t = T_t + S_t + \varepsilon_t, \quad t = 1,\ldots,n,
\end{equation}
where $T_t$ represents the piecewise linear function  
$T_t = \alpha_j + \beta_j\,t$, $\tau_{j-1}< t\leq \tau_j$,
with abrupt changes at $\tau_0 < \tau_1 < \cdots < \tau_m$, $S_t$ describes
a seasonal component, $\varepsilon_t$ denotes white noise with constant variance $\sigma^2$,
and $n$ is the sample size. 
Abrupt changes $\tau_k$s are estimated using
\texttt{R} package \texttt{bfast}, cf.~\cite{verbesselt2012package}.
The user can provide the significance level of the
abrupt change test 
and must set the test statistic bandwidth $h$. 
For our applications we use a $5\%$ significance level.
More details on $h$ can be found in Section VIII-{\bf A}
in the Supplementary Materials.


Let $\wh{\tau}_k$ denote the $k$-th abrupt change estimated by BFAST 
in an NDVI time series.
The dNBR is an appropriate variable to assess wildfire severity in forest vegetation,
cf.~\cite{escuin2008fire}.
Thus, given $\wh{\tau}_k$, its corresponding dNBR 
is defined as
$\mbox{dNBR}(\wh{\tau}_k) = \mbox{NBR}_{\wh{\tau}_k-23} - \mbox{NBR}_{\wh{\tau}_k+1}$.
Note that the time-point $\wh{\tau}_k-23$ corresponds to almost a year earlier 
than the estimated breakpoint $\wh{\tau}_k$ whereas $\wh{\tau}_k+1$ is the time-point 
right after the estimated vegetation change.
Using these 2 dates minimizes the differences specifically linked to phenological 
changes or illumination conditions, cf.~\cite{escuin2008fire}.

Having calculated $\mbox{dNBR}(\wh{\tau}_k)$ we use the values of
Table~\ref{tab_burnSeverity}, adapted from \cite{key2006landscape},
to classify the vegetation change type near $\wh{\tau}_k$. 
For burned area determination
we consider 2 categories: unburned ($\mbox{dNBR} < 0.1$) and burned ($\mbox{dNBR} \geq 0.1$).
It is customary to summarize the abrupt change estimated by BFAST
to the year level, see Section~4.2 of \cite{verbesselt2010detecting}.
Thus, we produce annual burned and severity burn maps.

\begin{table}[hbt]
\caption{\small Vegetation change types.}~\label{tab_burnSeverity}
\centering
\scalebox{0.65}{
\begin{tabular}{ccc}
\toprule[1.25pt]
    {\bf dNBR} & Regrowth & Severity \\
    
	\midrule[1.25pt]  	
	
	        $< -0.25$ & High\\
	$-0.25$ to $-0.1$ & Low\\
      $-0.1$ to $0.1$ & & Unburned\\
      $0.1$ to $0.27$ & & Low\\
     $0.27$ to $0.66$ & & Moderate\\
     $> 0.66$ & & High\\      	        
  \bottomrule[1.25pt]
\end{tabular}
}
\end{table}


\section{Data sets}

In order to calculate 
NDVI and NBR we only need spectral 
bands 3, 4, 5 and 7 from the Landsat-7 images. After proper processing
at the L1T level and cloud masking we 
had 238 of the 322 images expected in the studied period. 
From our simulations (see Section VIII in the Sup. Mate.), we know
that combining linear interpolation and BFAST for estimating
one abrupt change in time series with at least $30\%$ missing 
values yields marginally better results than using spline interpolation.
Therefore,
we considered linear and spline interpolation as gap filling 
methods for the NDVI and NBR stack of images.
For comparison purposes, in our application we also
applied the ArcMap~10.3.1's low-pass filter with a $3\times 3$ spatial kernel smoother,
as well as the algorithm 
\texttt{gdal\_fillnodata} from the GDAL library, cf.~\cite{GDAL}
to the 238 Landsat-7 images; all remaining gaps were filled in using linear interpolation.

To validate our approach
we used a RapidEye burned area polygon for 2012.
This polygon's estimated burned area is based on NDVI, the 
burned area index proposed by \cite{martin2001propuesta},
and some empirical thresholds 
(chosen via visual inspection).
This polygon can be seen in Figure~\ref{FIG:1A}
along with others provided by governmental agencies.

\begin{figure}[hbt]
    \centering
        \scalebox{1}{\includegraphics[width=0.75\linewidth]{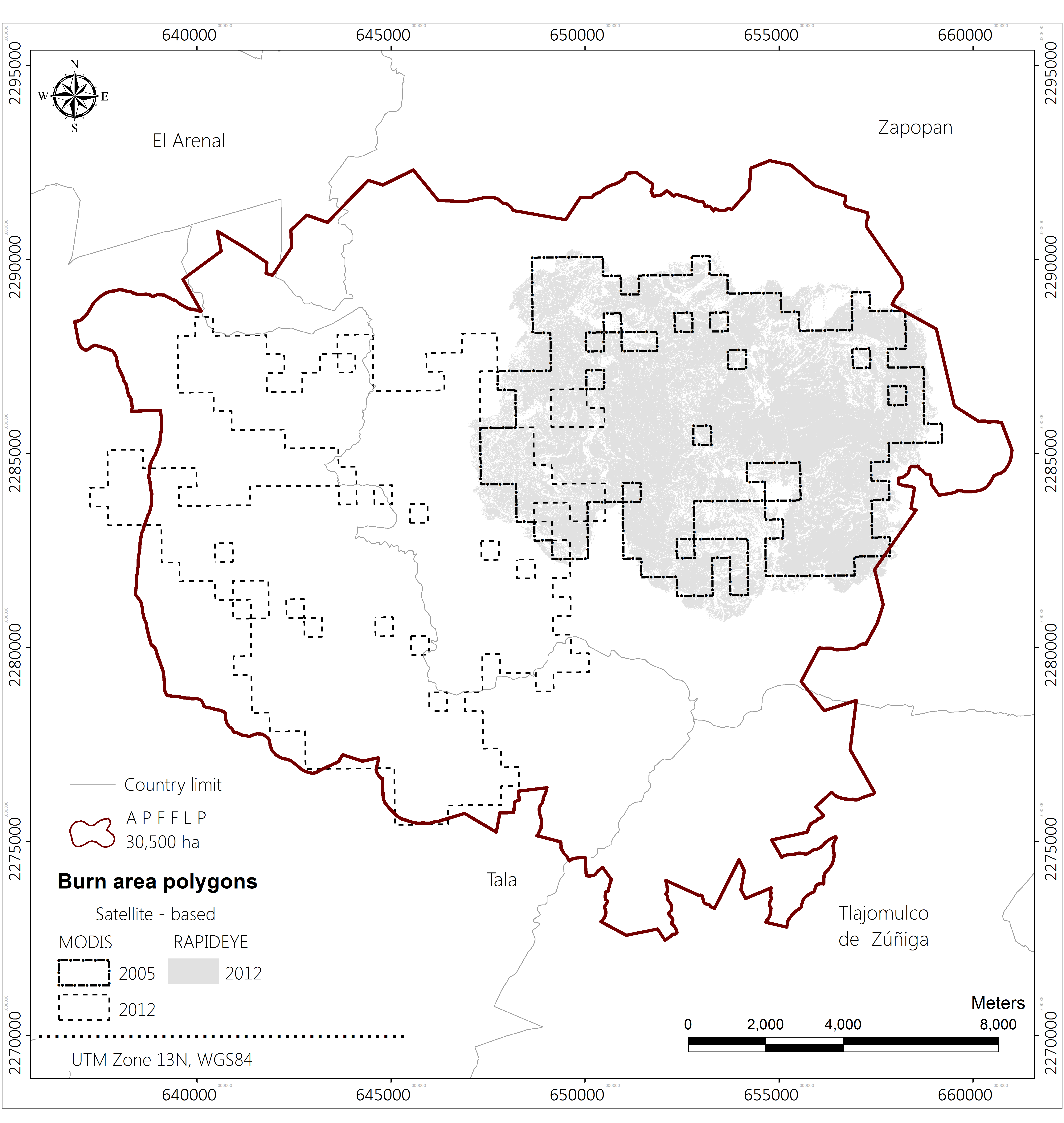}}\\
        \scalebox{1}{\includegraphics[scale=0.15]{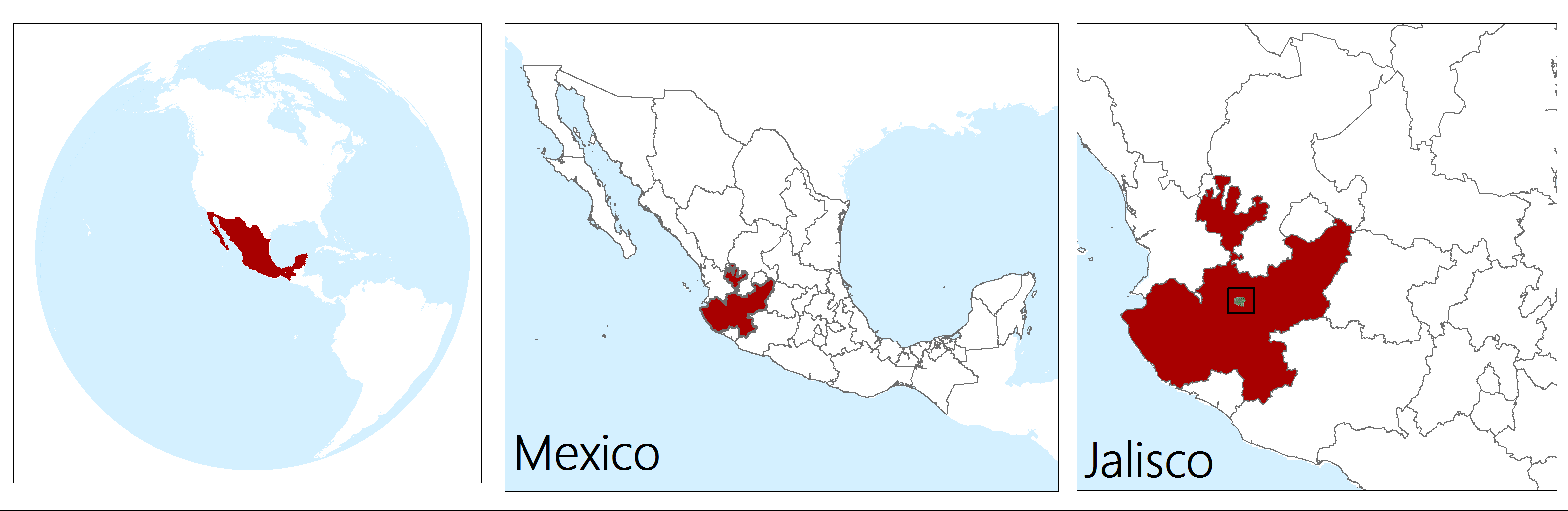}}%
	\caption{\small Polygons of estimated burned areas based on MODIS and RapidEye images in La Primavera.}
	\label{FIG:1A}
\end{figure}

\section{Results}
%
Reports of wildfires in La Primavera are available for the period 1998-2012,
cf.~\cite{huerta2014incendios}. 
Our method detected burned areas
between 2005 and 2014, see Table~\ref{tab_burnAreaBySeverity}.
Below we comment on the validation
and burn severity of some of these burned area maps.

\subsection{Accuracy assessment of burned area maps}~\label{sec.accuracy.assessment}

\begin{figure*}[hbt]
    \centering 
	\includegraphics[scale=.08]{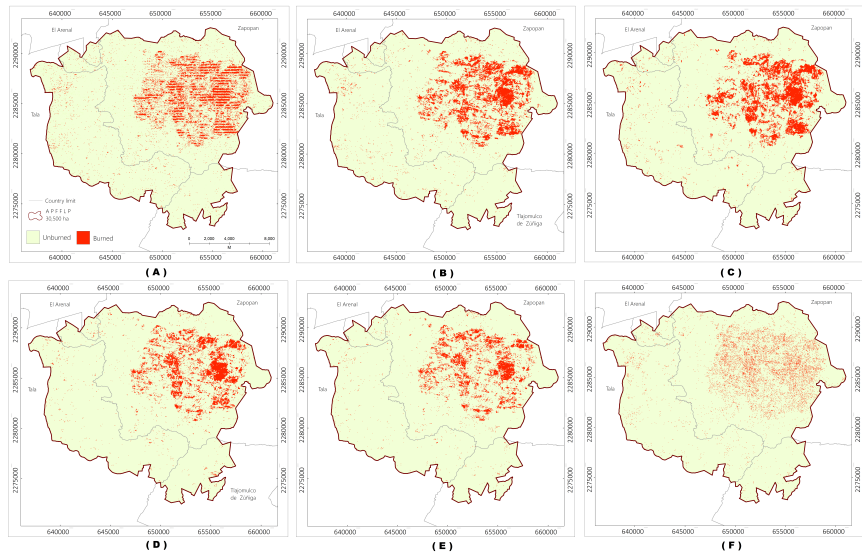}
    \caption{\footnotesize {2012 burned area maps. 	        
    ({\bf A}) Lineal interpolation and $h=0.15$. 
    ({\bf B}) Lineal interpolation and $h=0.23$.
    ({\bf C}) Spline interpolation and $h=0.15$.
    ({\bf D}) Spline interpolation and $h=0.23$.
    ({\bf E}) Interpolated with \texttt{gdal\_fillnodata} and $h=0.15$.
    ({\bf F}) Interpolated with ArcMap and $h=0.23$.
    }}
    \label{FIG:8A}
\end{figure*}

In this section we 
assess the 
overall accuracy of our 2012 burned area map 
based on Ch.~4 of \cite{congalton2002assessing} (see Figure~\ref{FIG:8A}). 
Our aim is to assess 
overall accuracy as a function of data quality.
To this end, 
we measure the quality of
each pixel 
based on
the percentage of missing data in the 
corresponding NDVI time series
(see Fig.~5 in the Sup.~Materials).
A pixel is said 
to have \emph{poor quality} if 
it has
$50$ to $53\%$ missing data. 
Similarly, a \emph{moderate quality} pixel
is missing
$47$ to $49\%$ of 
data. 
Based on
these definitions,
the
corresponding overall accuracy 
was computed (see Table~\ref{tab_overallAccuracy}).

\begin{table*}[hbt]
\caption{\footnotesize Overall accuracy of 2012 burned area map as a function
of data quality and gap filling strategies.}~\label{tab_overallAccuracy}
\centering
\scalebox{0.6}{
\begin{tabular}{cccccc}
\toprule[1.25pt]    	
    \multicolumn{2}{c}{Gap filling method}
    &&
    \multicolumn{3}{c}{Overall accuracy ($\%$)}\\
    \cmidrule[1.25pt]{1-2} \cmidrule[1.25pt]{4-6}        

	{\bf Spatial} & {\bf Temporal} & {\bf h} & Whole area & Poor data quality & Moderate data quality \\
	\midrule[1.25pt]        
	
			& \mbox{Linear} & $0.15$ & 59.62 & 70.26 & 92.36 \\
			& \mbox{Linear} & $0.23$ & 58.93 & 69.63 & 90.71 \\
			& \mbox{Spline} & $0.15$ & 61.75 & 70.52 & 91.61 \\
			& \mbox{Spline} & $0.23$ & 56.76 & 67.03 & 90.39 \\
	\texttt{gdal\_fillnodata} & \mbox{Linear} & $0.15$ & 63.31 & 72.74 & 92.61 \\			
	     \mbox{ArcMap filter} & \mbox{Linear} & $0.23$ & 51.05 & 64.28 & 90.25 \\
        
  \bottomrule[1.25pt]
\end{tabular}
}
\end{table*}

If all the pixels of the reference RapidEye polygon are considered (see  
\emph{Whole area} in Table~\ref{tab_overallAccuracy}), 
the overall accuracy of our maps is rather low, $51$ to $63\%$. 
In contrast, if we focus only on areas with poor quality pixels, this
overall accuracy improves $11\%$ (on average).
This improvement raises even further ($33\%$)
when we consider areas with moderate quality pixels. 
This progressive improvement (from $58.57$ to $91.32\%$ on average) in the overall 
accuracy suggests that our approach will produce appropriate burned area maps
in scenarios where 
a small amount of data are missing in the original data cubes (NDVI and NBR).

We also evaluate the effect of BFAST's bandwidth parameter $h$
on our 2005 and 2008 burned area maps.
Of the six plots in Figure~\ref{FIG:8A}, {\bf A} and {\bf F} show 
a clear spatial discontinuity in the estimated
burned area. 
This apparent discontinuity
seems to be resolved when the BFAST's bandwidth increases ($h=0.23$) and when we use
splines as the temporal interpolation method ({\bf C}-{\bf E}). 
Setting 
these visual features aside, the maps in which $h=0.15$ show the highest 
overall accuracy irrespective of data quality and gap-filling method. 
For instance, with \emph{Moderate data quality}
the 
spatial-temporal gap filling approach (Figure~{\bf D} and \texttt{gdal\_fillnodata} row
in Table~\ref{tab_overallAccuracy}) seems to produce the same effects on overall accuracy
as when only a temporal linear interpolation is applied (Figure~{\bf A}).

\subsection{Burned area maps for La Primavera with severity assessment}
%
Burn severity provides a measure of the damage
inflicted by fire on an ecosystem, cf.~\cite{key2006landscape}.
In order to quantify burn severity, we use
Table~\ref{tab_burnSeverity}.
\begin{table*}[htb]
\caption{\small Estimated burned hectares by year and
fraction of burned hectares in La Primavera by level of severity.
Here \texttt{NA} stands for \emph{not available}.}~\label{tab_burnAreaBySeverity}
\centering
\scalebox{0.6}{
\begin{tabular}{c cc c cc c cc c cc}
\toprule[1.25pt]
    	
    & \multicolumn{2}{c}{\bf Total burned area (ha)} 
    && \multicolumn{8}{c}{\bf Severity Level}\\
    \cmidrule[1.25pt]{2-3}
    \cmidrule[1.25pt]{5-12}   
    
    &&  
    && \multicolumn{2}{c}{\bf Low} 
    && \multicolumn{2}{c}{\bf Moderate} 
    && \multicolumn{2}{c}{\bf High}\\
    \cmidrule[1.25pt]{5-6} \cmidrule[1.25pt]{8-9} \cmidrule[1.25pt]{11-12}        

	{\bf Year} & $h=0.15$ & $h=0.23$ &&
	             $h=0.15$ & $h=0.23$ && 
	             $h=0.15$ & $h=0.23$ && 
	             $h=0.15$ & $h=0.23$\\
	\midrule[1.25pt]        
	
	2005  & 3549.24 & \texttt{NA} && 0.648 & \texttt{NA} && 0.349 & \texttt{NA} && 0.003 & \texttt{NA} \\
	2006  &  189.27 &      729.33 && 0.888 &       0.775 && 0.108 &       0.224 && 0.004 & 0.001 \\
	2007  &  437.58 &      482.01 && 0.922 &       0.933 && 0.077 &       0.067 && 0.001 & 0.000 \\
	2008  & 1793.34 &     1156.78 && 0.695 &       0.538 && 0.292 &       0.431 && 0.014 & 0.031 \\ 
	2009  &  499.86 &      304.45 && 0.874 &       0.951 && 0.126 &       0.049 && 0.000 & 0.000 \\ 
	2010  &  402.84 &      327.18 && 0.916 &       0.870 && 0.083 &       0.128 && 0.001 & 0.003 \\
	2011  &  645.48 &      620.86 && 0.908 &       0.898 && 0.092 &       0.102 && 0.000 & 0.000 \\
	2012  & 2594.97 &     2180.61 && 0.590 &       0.810 && 0.398 &       0.185 && 0.012 & 0.005 \\
	2013  &  839.16 &     1596.44 && 0.859 &       0.871 && 0.139 &       0.128 && 0.002 & 0.001 \\
	2014  &  454.95 &  \texttt{NA}&& 0.938 & \texttt{NA} && 0.062 & \texttt{NA} && 0.000 & \texttt{NA} \\
	        
  \bottomrule[1.25pt]
\end{tabular}
}
\end{table*}
\begin{figure*}[htb]
    \centering
	\includegraphics[scale=.08]{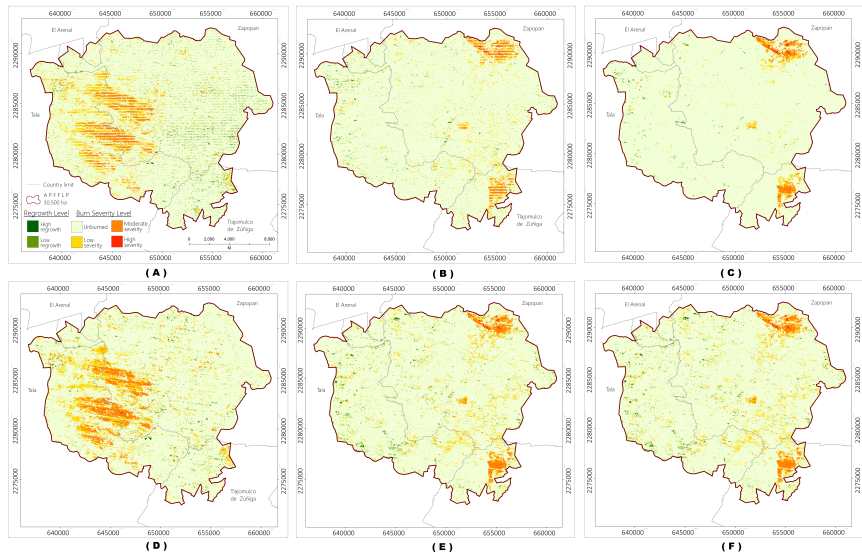}
    \caption{\footnotesize {Burn severity maps for 2005 and 2008.
    Linear interpolation.
    ({\bf A}) 2005, $h=0.15$. 
    ({\bf B}) 2008, $h=0.15$.
    ({\bf C}) 2008, $h=0.23$.
    Spline interpolation. 
    ({\bf D}) 2005, $h=0.15$.        
    ({\bf E}) 2008, $h=0.15$.
    ({\bf F}) 2008, $h=0.23$.
    }}
    \label{FIG:8}
\end{figure*}
Table~\ref{tab_burnAreaBySeverity}
shows that La Primavera's burn severity is rather low between 2005 and 2014.
Because high burn severity has been shown to cause
high erosion rates and low vegetation recovery,
cf.~\cite{doerr2006effects} and \cite{moody2013current}, 
we can infer that most of La Primavera's burned areas suffer from  
low erosion and profit from a high vegetation regrowth rate.

\begin{figure*}[htb]
    \centering
        \scalebox{1}{\includegraphics[width=0.25\linewidth]{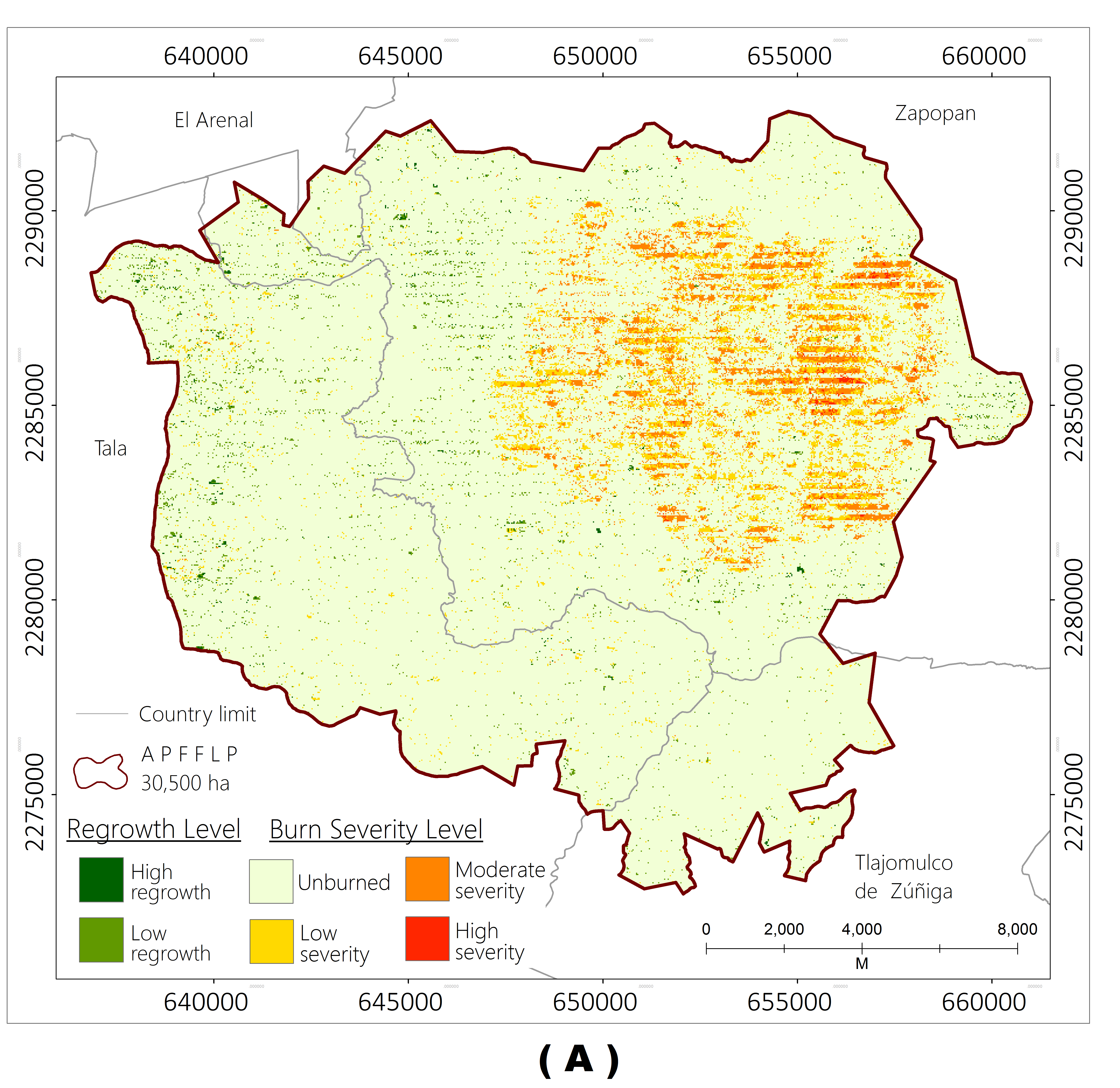}}%
        \scalebox{1}{\includegraphics[width=0.25\linewidth]{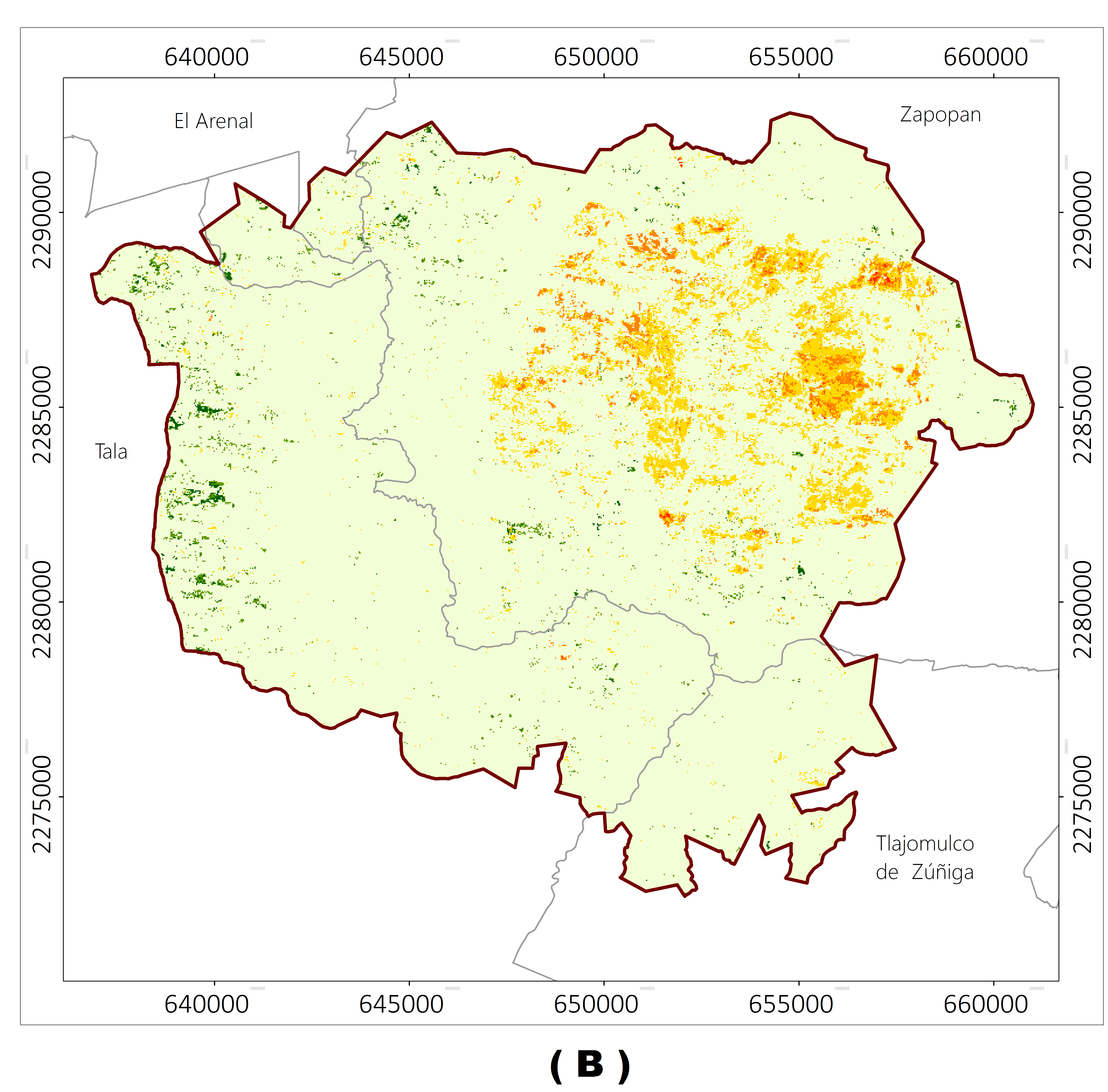}}%
        \scalebox{1}{\includegraphics[width=0.25\linewidth]{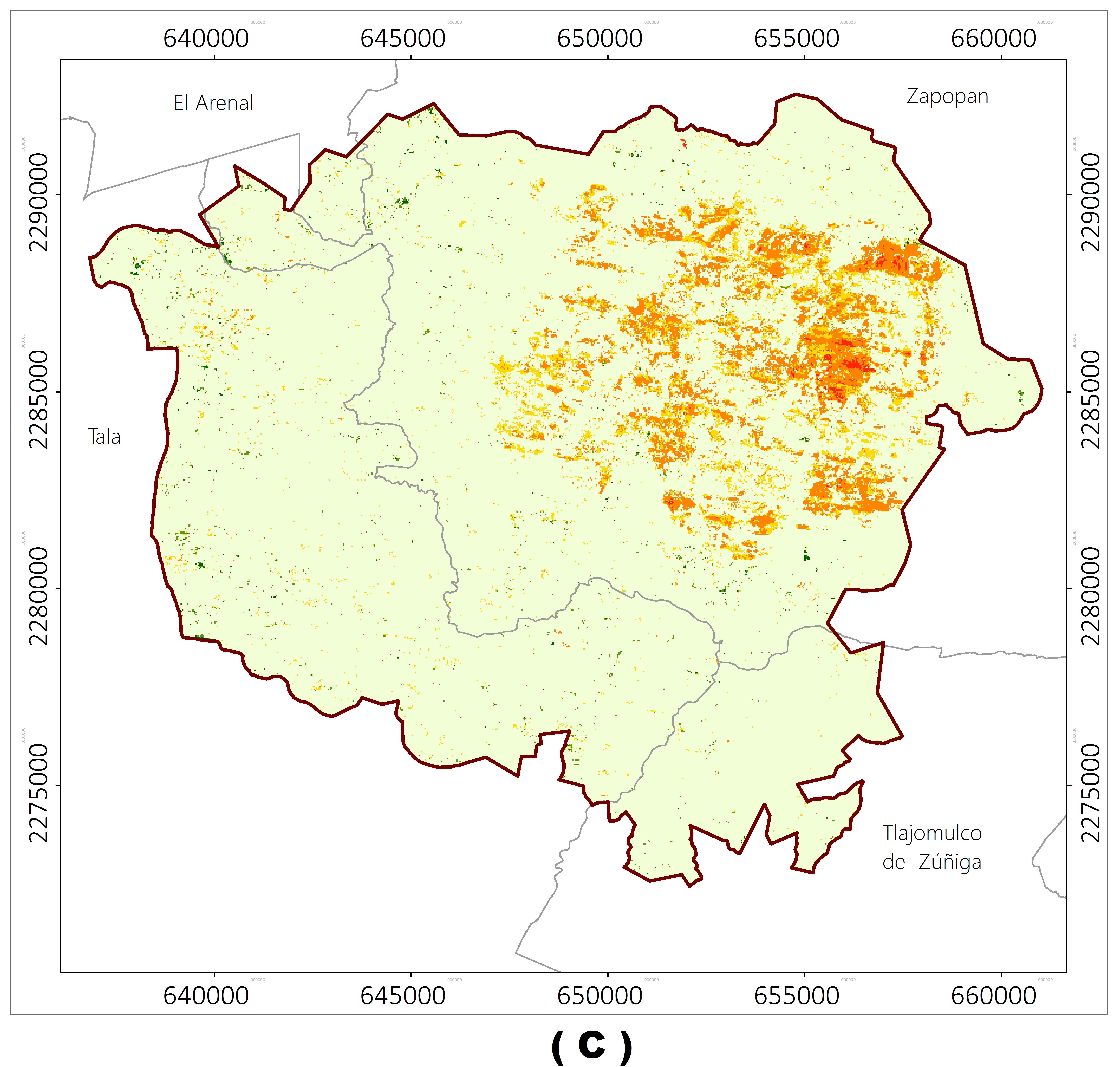}}%
		\scalebox{1}{\includegraphics[width=0.25\linewidth]{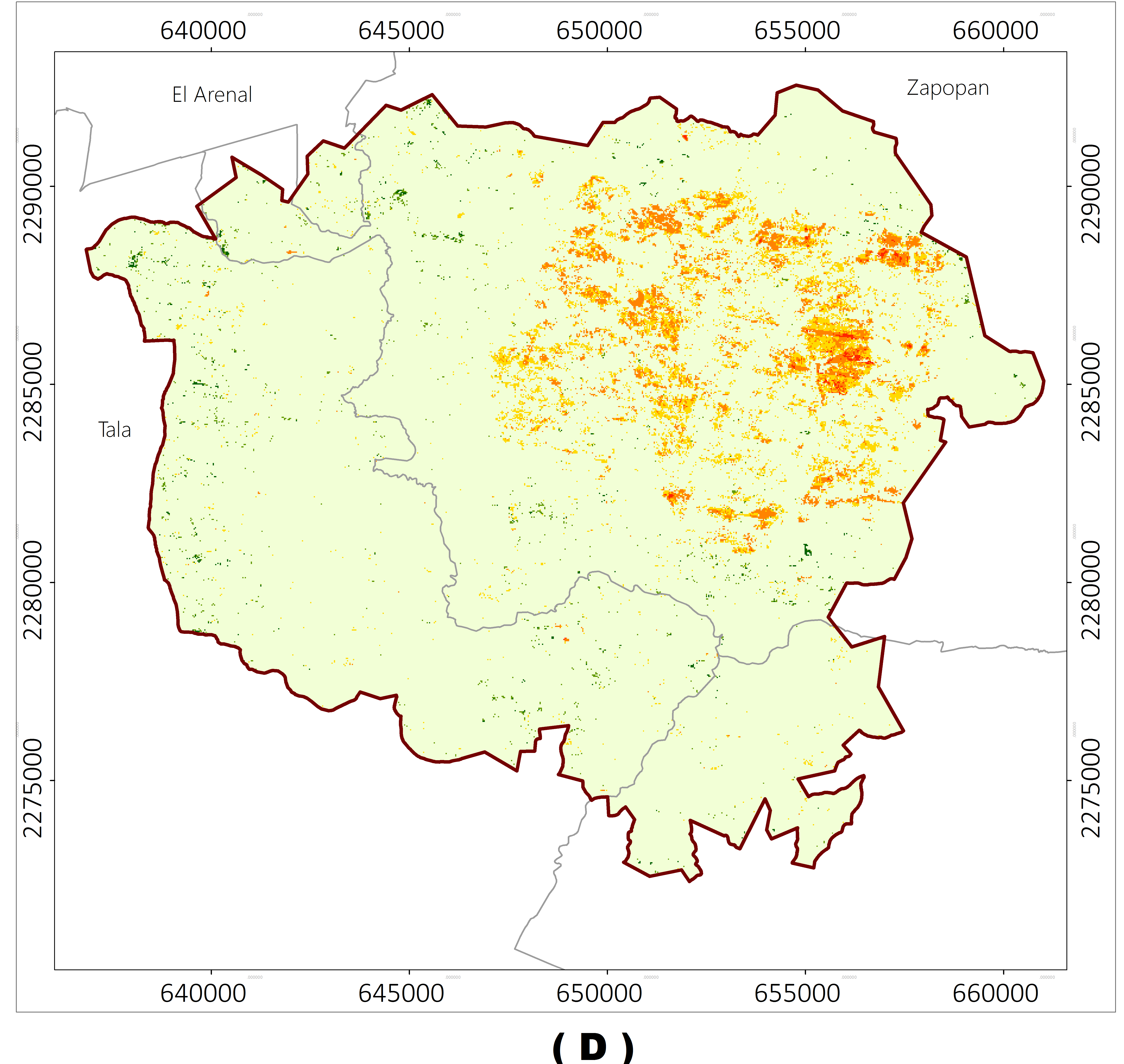}}%
    \caption{\footnotesize {Burn severity maps for 2012.
    Linear interpolation.
    ({\bf A}) $h=0.15$. 
    ({\bf B}) $h=0.23$.
    Spline interpolation.
    ({\bf C}) 2012, $h=0.15$. 
    ({\bf D}) 2012, $h=0.23$.        
    }}
    \label{FIG:9}
\end{figure*}

We now comment on our burn severity maps for the 2005, 2008
and 2012 events.
Figure~\ref{FIG:8}-{\bf A} and {\bf D} report on the 2005 burned area.
In both figures, we require the use of $h=0.15$ as BFAST's bandwidth
parameter (see Section~VIII in the Sup.~Mate.~for further details).
In these maps, most of the pixels have been classified as low severity
followed by moderate, and to a far less degree, high severity.
In 2005, the affected vegetation was primarily native and induced grasslands,
followed by shrubs and bushes, and to a less extent, adult woodlands (oak-pine forests).
Figure~\ref{FIG:8}-{\bf B, C, E} and {\bf F} show the 2008 burned area 
in the
Northeast and Southeast of La Primavera. 
The majority
of this burned area is categorized as low severity, regardless of the bandwidth
value. 
The fraction of burned area categorized as moderate severity
is roughly 1.5 times larger for $h=0.23$ than for $h=0.15$. 
Also, the fraction of burned area categorized under high severity
is twice as larger for $h=0.23$ than for $h=0.15$. 
The vegetation of this area includes oak-pine forests and
induced grasslands; to this date there is no official record
of the total area damaged
(see \cite{huerta2014incendios} for more details). 
Figure~\ref{FIG:9} shows the 2012 burned area. 
Independently of the bandwidth value,
most of the estimated burned area has been categorized as low severity.
Differences seem to emerge for the moderate and high severity categories;
the fraction of burned area categorized as moderate severity
is twice as larger for $h=0.15$ than for $h=0.23$; a similar result holds
for the high severity category.
Oak-pine forests and shrubby vegetation dominate this area. According
to 
the state of Jalisco, 10 to 20\% of wooded area was lost
as a result
of this fire, which might have been started in a clandestine landfill.

\section{Conclusion}~\label{sec.conclusion}
The proposed method is a semi-automatic approach for burned area mapping
and, subsequently, burn severity assessment. 
Moreover, unlike others, this method provides burned area maps with statistical guarantees.
Additionally, 
using annual burn severity maps,
the user 
can access
information 
on
vegetation recovery 
that may prove
useful in many subsequent studies, for instance in fire ecology.
The overall accuracy of the 
annual burned area maps
was evaluated in response to the existence of
differing amounts of missing data.
The overall accuracy of the burned area map
that we were able to validate 
increased from a modest
69\% (pixels with poor quality data) to a remarkable 92\% (pixels with 
moderate data quality). This provides evidence that
BFAST can be used as an effective and easy-to-apply
method for burned area mapping.

\section{Acknowledgements}~\label{sec.acknow}

{This research was funded by CONACyT's 
Convocatoria de Proyectos de Desarrollo Cient\'ifico 
para Atender Problemas Nacionales 2016 grant number 2760.}
{The authors would like to thank Lilia de Lourdes Manzo Delgado, 
Leticia G\'omez Mendoza and St\'ephane Couturier with the Institute of Geography
at UNAM as well as Rainer Ressl with CONABIO for helpful discussions and suggestions.
Special thanks to Roberto Mart\'inez with CONABIO for pointing out the function
\texttt{gdal\_fillnodata}.}

\bibliographystyle{ieeetr} 
\bibliography{proxyBurnBib}

\newpage
{\bf Supplementary Materials}
\section{Supplementary Materials}~\label{sec.supp}
Figure~\ref{FIG:1A} shows the percentage of missing values in the 
Landsat-7 NDVI images taken from 2003 to 2016 over
La Primavera Flora and Fauna Protection Area in Jalisco, Mexico.
In the main manuscript of this letter we used a free-gap version of this
data cube to show BFAST's potential as a tool for mapping burned
areas.
In this Supplementary Materials we
evaluate BFAST's performance to estimate abrupt changes in
synthetic time series sharing the main features, including missing values,
of La Primavera's Landsat-7 NDVI images. 
We use linear and spline interpolation to fill the
gaps of the synthetic time series.

We simulate 16-day NDVI time series from 2003 to 2016 obeying the additive
representation given by Eq.~$(1)$, see main manuscript. The seasonality ($S_t$) and 
the white noise ($\varepsilon_t$) of these synthetic time series are simulated with a harmonic
regression model and samples of a normal distribution with zero mean and
standard deviation ($\sigma$), respectively.
In the next sections we will specify the trend structure, $T_t$, as this is
different in each study.
Each simulation study was repeated 1000 times and we use metrics
such as probability coverage and mean squared error (MSE) to assess 
BFAST's performance.

\begin{figure}[hbt]
    \centering
        \scalebox{1}{\includegraphics[width=0.75\linewidth]{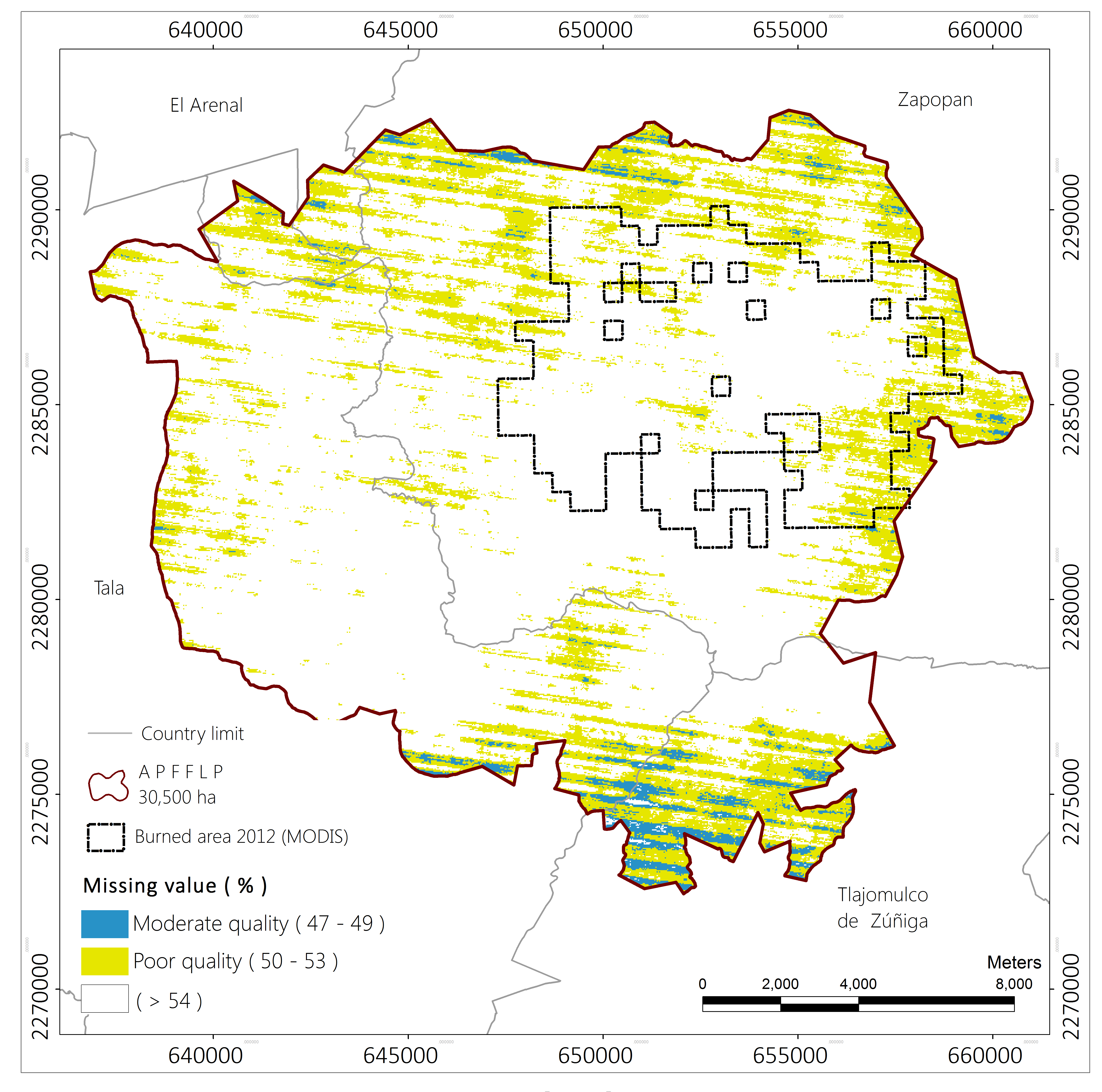}}\\
        \scalebox{1}{\includegraphics[scale=0.15]{LaPrimaveraLocation.png}}%
	\caption{\small Overview of data quality of data cubes for La Primavera: Percentage
	of missing values in each NDVI pixel.}
	\label{FIG:1A}
\end{figure}

\section{Simulations}~\label{sec.sims}
For the amplitude parameter needed in the harmonic model we consider 
$0.15, 0.3$ and $0.45$; for simplicity we use 
a $0^\circ$ phase angle.
We consider $\sigma = (0.02, 0.05, 0.07)$ as these values 
are in line with the variability level found in La Primavera's 
Landsat-7 NDVI time series.

We \emph{simulate} missing values as follows. 
From the 322 time-points of any
simulated time series ($y_t$) we choose randomly and without replacement, $t_1, \ldots, t_P$ 
say, time-points. Observe that these $P$ time-points are chosen at random, and
consequently, there is no order relation among them.
Then, the corresponding value in the time series is masked as \emph{not
available}, that is, we set $y_{t_i}=\texttt{NA}$, where $i=1,\ldots,P$. In 
the simulations below we use $P = 0, 32, 64, 97, 129, 161, 194$. 
Approximately, these values corresponds to
$0, 10, \ldots, 50$ and 60\% of the total 322 observations.

\subsection{On BFAST's bandwidth parameter}

BFAST utilizes the OLS-MOSUM test to determine statistically 
the existence of abrupt changes. 
This test is based on a sequence of partial sums of ordinary least-squares 
residuals, cf.~\cite{Chu.etal.1995} and \cite{Zeileis.etal.2002};
the number of residuals in each sum is fixed, approximately $nh$, but
controlled by a \emph{bandwidth} parameter $0 < h < 1$; $n$ denotes sample size. 
Although there are 
some empirical rules to select the value of $h$, cf.~Section 2.2 of \cite{verbesselt2010phenological}, 
our data set does not meet the conditions for these rules to be applied, specially
due to the large amount of missing data. Hence, we also include $h$ as a parameter
in our simulations; we select $h = 0.15, 0.23, 0.45$.  
For a time series of 322 observations, $h=0.45$
is equivalent to using a bandwidth of 145 time-points in the
aforementioned OLS-MOSUM statistic; this amount of time-points is 
equivalent to 
6 years, approximately, in the Landsat temporal scale. 

\subsection{Assessing BFAST's performance in estimating one abrupt change}

In this study the trend has a single abrupt change at the observation 161.
Before this observation the trend is constant ($0.7$)
and afterwards follows the line $0.3 + 0.2/161\,t$. More precisely and following
Eq.~(1), $n=322$, $\alpha_1=0.7$, $\alpha_2=0.3$, $\beta_1=0$, 
$\beta_2=0.2/(161)$, $\tau_0=0$, $\tau_1=161$, $\tau_2=322$ and $m=2$.

As a first goal we are interested in the \emph{probability coverage} of BFAST
to estimate
one abrupt change, i.e., the number of times in which an abrupt change 
is detected divided by the number of simulations.
We study probability coverage
as a function of the s.d.~of the errors of model~(1), interpolation
method, percentage of missing values, amplitude of the seasonal component
and the bandwidth value $h$ utilized by BFAST.

Since a small s.d.~benefits both interpolation methods, 
analysis not included, 
here we present the results of simulations when 
we set $\sigma=0.02$ (the smaller value of $\sigma$ under consideration) 
and allow the bandwidth parameter to vary.

Let us discuss the results of Table~\ref{tab_metodos_missVal_h}.
Observe that the probability coverage is greater than $0.9$ independently 
of amplitude, bandwidth and interpolation method even when 20\% of observations
are missing.

\begin{table}[htb]
\caption{\footnotesize Probability ($\times 100$) coverage of estimating one abrupt change as
a function of the amplitude of the simulated time series,
the BFAST's bandwidth $h$ and the percentage of missing values;
here $\sigma = 0.02$.}\label{tab_metodos_missVal_h}
\centering
\scalebox{0.55}{
\begin{tabular}{cccccccccc} 
\toprule[1.25pt] 
	
    &&& \multicolumn{7}{c}{\% of missing values}\\
    \cmidrule[1.25pt]{4-10}
        
    {\bf amplitude} & {\bf h} & {\bf methods} & 0 & 10 & 20 & 30 & 40 & 50 & 60 \\ 
    \midrule[1.25pt]     
    
    $0.15$ &
    $0.15$ &
    Linear & 99.2 & 97.9 & 92.7 & 79.0 & 55.9 & 29.1 & 10.9 \\ 
    &&
    Spline & 99.2 & 98.2 & 92.4 & 83.1 & 68.9 & 55.6 & 41.5 \\
    
    & $0.23$ &
    Linear & 99.9 & 99.7 & 97.6 & 92.1 & 80.5 & 60.3 & 39.9 \\ 
    &&
    Spline & 99.9 & 99.3 & 97.4 & 93.6 & 88.3 & 79.8 & 68.6 \\
    
    & $0.45$ &
    Linear & 100 & 100 & 100 & 100 & 100 & 100 & 100 \\ 
    &&
    Spline & 100 & 100 & 100 & 100 & 100 & 100 & 100 \\
    
    \midrule[1.25pt]
    
    $0.3$ &
    $0.15$ &
    Linear & 99.8 & 98.9 & 95.1 & 79.7 & 55.4 & 23.6 & 7.9 \\ 
    &&
    Spline & 99.8 & 99.1 & 95.6 & 88.7 & 79.4 & 67.5 & 53.3 \\ 
    
    & $0.23$ &  
  	Linear & 100 & 99.9 & 98.4 & 92.4 & 78.1 & 56.1 & 35.1 \\ 
  	&&
	Spline & 100 & 99.8 & 99.0 & 96.0 & 91.9 & 85.4 & 74.9 \\ 
	
	& $0.45$ &
	Linear & 100 & 100 & 100 & 100 & 100 & 100 & 100 \\ 
	&&
    Spline & 100 & 100 & 100 & 100 & 100 & 100 & 99.9 \\ 
    
	\midrule[1.25pt]
	
	$0.45$ &
	$0.15$ &
	Linear & 100 & 99.4 & 95.7 & 80.6 & 54.9 & 22.6 & 6.9 \\ 
	&&
	Spline & 100 & 99.5 & 97.6 & 94.3 & 87.6 & 75.9 & 59.8 \\ 
	
	& $0.23$ &		
  	Linear & 100 & 99.9 & 98.6 & 92.6 & 76.7 & 53.4 & 31.8 \\ 
  	&&
	Spline & 100 & 99.9 & 99.4 & 98.3 & 95.4 & 89.4 & 78.8 \\ 
	
	& $0.45$ &
  	Linear & 100 & 100 & 100 & 100 & 100 & 100 & 100 \\ 
  	&&
    Spline & 100 & 100 & 100 & 100 & 100 & 100 & 99.5 \\ 
                                          
  \bottomrule[1.25pt]
\end{tabular}
}
\end{table}

The probability coverage is remarkably large when $h=0.45$ independently
of amplitude and percentage of missing values. 
From our next simulation study (Section~\ref{sec.sims.2CPS})
we infer that utilizing $h=0.45$
in our real data application is equivalent to requiring that
the separation between two abrupt changes be of roughly 6 years.
Due to this and because our a priori information reports vegetation changes 
in 2005, 2010, 2012 and 2013, we will not use $h=0.45$ in our application.

For $h = 0.15, 0.23$ the probability coverage decreases
as the percentage of missing values increases. Note that when we use
spline-based interpolation the probability coverage is at least
$0.68$ in the difficult case of having 60\% missing values.
In this regard, spline is clearly better than linear interpolation.
Observe, however, that this feature only means that is more likely
to estimate \emph{one abrupt change} when we use spline than when we use
linear interpolation. This does not tell us much about estimating the
\emph{correct abrupt change} though. Assessing this characteristic is the goal
of our next simualation.

\begin{table}[htb]
\caption{\footnotesize Correct estimation probability coverage ($\times 100$) 
of estimating $\wh{\tau}_1=161$ as a function of the amplitude of 
the simulated time series, the BFAST's bandwidth $h$ and the percentage 
of missing values; here $\sigma = 0.02$.}\label{tab_metodos_missVal_h_rightEstimation}
\centering
\scalebox{0.55}{
\begin{tabular}{cccccccccc} 
\toprule[1.25pt] 
	
    &&& \multicolumn{7}{c}{\% of missing values}\\
    \cmidrule[1.25pt]{4-10}
        
    {\bf amplitude} & {\bf h} & {\bf methods} & 0 & 10 & 20 & 30 & 40 & 50 & 60 \\ 
    \midrule[1.25pt]     
    
    $0.15$ &
    $0.15$ &
    Linear & 100 & 91.7 & 80.5 & 69.7 & 58.9 & 53.6 & 42.2 \\ 
    &&
    Spline & 100 & 91.4 & 80.2 & 70.2 & 60.8 & 47.8 & 37.1 \\
    
    & $0.23$ &
    Linear & 100 & 91.4 & 80.6 & 70.6 & 60.2 & 53.4 & 45.1 \\  
    &&
    Spline & 100 & 91.2 & 80.1 & 69.9 & 60.9 & 49.6 & 36.3 \\
    
    & $0.45$ &
    Linear & 100 & 91.4 & 80.5 & 70.2 & 60.3 & 52.6 & 42.4 \\
    &&
    Spline & 100 & 91.3 & 80.1 & 70.0 & 61.7 & 50.0 & 37.0 \\
    
    \midrule[1.25pt]
    
    $0.3$ &
    $0.15$ &
    Linear & 100 & 91.3 & 80.8 & 69.4 & 59.0 & 55.1 & 43.0 \\  
    &&
    Spline & 100 & 91.3 & 80.4 & 69.2 & 61.5 & 49.0 & 35.3 \\ 
    
    & $0.23$ &  
  	Linear & 100 & 91.4 & 80.7 & 70.6 & 61.7 & 53.8 & 45.3 \\  
  	&&
	Spline & 100 & 91.3 & 79.9 & 69.5 & 62.0 & 50.8 & 35.0 \\
	
	& $0.45$ &
	Linear & 100 & 91.4 & 80.5 & 70.9 & 62.3 & 54.8 & 44.8 \\  
	&&
    Spline & 100 & 91.3 & 80.0 & 70.0 & 61.7 & 50.8 & 35.2 \\ 
    
	\midrule[1.25pt]
	
	$0.45$ &
	$0.15$ &
	Linear & 100 & 91.3 & 81.1 & 69.9 & 58.7 & 51.8 & 50.7 \\ 
	&&
	Spline & 100 & 91.1 & 79.7 & 68.9 & 59.7 & 46.5 & 28.6 \\ 
	
	& $0.23$ &		
  	Linear & 100 & 91.4 & 81.0 & 71.1 & 62.1 & 56.6 & 44.7 \\ 
  	&&
	Spline & 100 & 91.0 & 79.2 & 69.0 & 59.7 & 47.2 & 30.1 \\ 
	
	& $0.45$ &
  	Linear & 100 & 91.4 & 80.9 & 71.6 & 64.0 & 56.7 & 45.1 \\ 
  	&&
    Spline & 100 & 91.0 & 79.3 & 69.5 & 59.8 & 47.0 & 31.0 \\ 
                                          
  \bottomrule[1.25pt]
\end{tabular}
}
\end{table}

We also assess BFAST's \emph{correct estimation} probability coverage, that is,
conditioned on having estimated an abrupt change, we computed the number
of times in which the BFAST estimate coincides with the true breakpoint, $\tau_1=161$;
here we allow amplitude and $h$ to vary but $\sigma = 0.02$, see Table~\ref{tab_metodos_missVal_h_rightEstimation}.
According to this table BFAST has a perfect correct estimation probability
coverage for one abrupt change when the time series is complete, that is,
\emph{does not} have missing values. 
As the amount of missing values in a time
series increases then the correct estimation deteriorates. Moreover, from this table we can
infer the following result. Let $0<p<1$ be given. When $100 \times p\%$ of observations 
are missing in a time series and BFAST has estimated one breakpoint,
then the probability that this is the true breakpoint is close to $1-p$. 
Additionally, when 50 to 60\% observations are missing from the simulated time series,
the correct estimation probability coverage is greater for linear interpolation
than for spline interpolation. This characteristic is more evident as the
amplitude increases.
This is relevant for our application as we deal with a fair amount of time 
series with roughly 50\% missing values.

\begin{figure}[hbt]
	\centering
		\includegraphics[scale=.325]{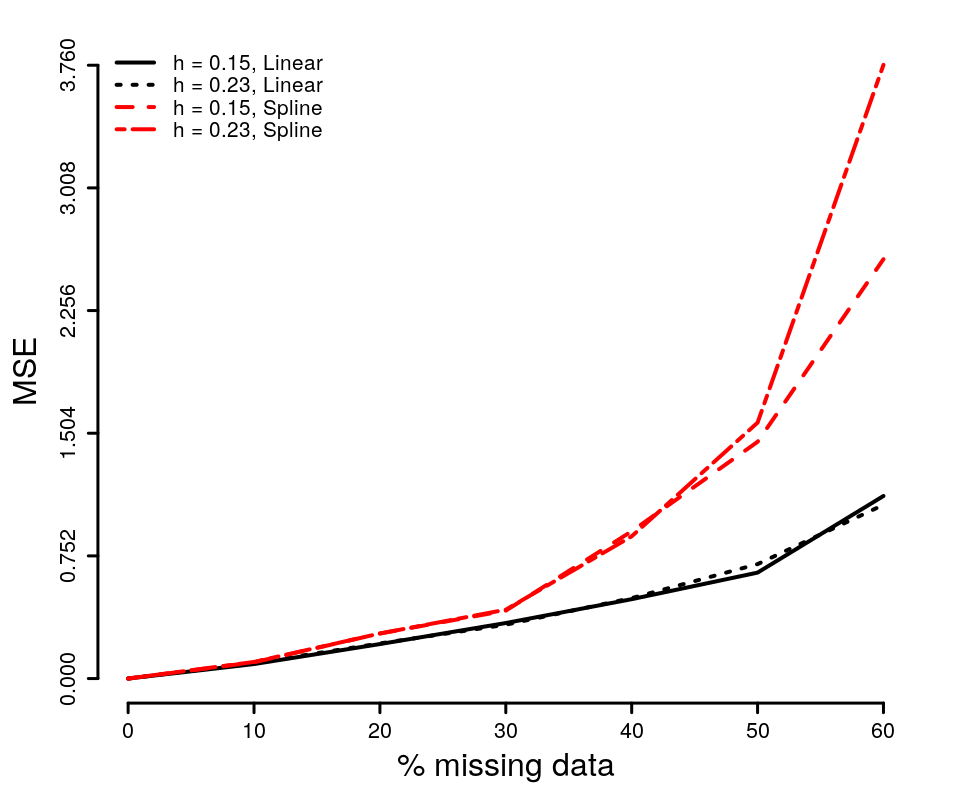}
	\caption{\footnotesize {BFAST's MSE for estimating an abrupt change with 
	$\sigma=0.02$ and amplitude $0.15$}}
	\label{FIG:2}
\end{figure}

Finally, BFAST's accuracy and precision to estimate an abrupt change
correctly is reported via the MSE.
Figure~\ref{FIG:2} shows that up to $20\%$ missing values,
BFAST's performance is appropriate independently 
of the parameter $h$ and interpolation method. From
$30\%$ and upwards, the combination of linear
interpolation and BFAST outperforms the combination of
spline interpolation and BFAST. 

\subsubsection{False negative identification}

It is also of interest to assess whether BFAST estimates an abrupt change in the
trend of time series with no such change. To this end, we follow Eq.~(1)
with $T_t \equiv 0$. As in the previous simulations, we allow the bandwidth value,
amplitude and percentage of missing values to vary. We set $\sigma=0.02$.

\begin{table}[hbt]
\caption{\footnotesize Probability ($\times 100$) of BFAST's false negative identification
as a function of the amplitude of the simulated time series, the BFAST's bandwidth 
$h$ and the percentage of missing values; here $\sigma = 0.02$.}\label{tab_metodos_missVal_h_falseNegative}
\centering
\scalebox{0.55}{
\begin{tabular}{cccccccccc} 
\toprule[1.25pt] 
	
    &&& \multicolumn{7}{c}{\% of missing values}\\
    \cmidrule[1.25pt]{4-10}
        
    {\bf amplitude} & {\bf h} & {\bf methods} & 0 & 10 & 20 & 30 & 40 & 50 & 60 \\ 
    \midrule[1.25pt]     
    
    $0.15$ &
    $0.15$ &
    
  Linear & 0.1 & 0.5 & 3.2 & 13.6 & 35.5 & 66.6 & 86.6 \\ 
  &&
  Spline & 0.1 & 0.8 & 3.9 & 11.4 & 24.6 & 39.0 & 53.9 \\ 
  & $0.23$ &
  Linear & 0.3 & 0.9 & 3.4 & 12.1 & 28.0 & 52.6 & 76.3 \\ 
  &&
  Spline & 0.3 & 0.6 & 3.7 & 8.4 & 17.2 & 30.5 & 43.2 \\ 
  & $0.45$ &
  Linear & 0.0 & 0.1 & 0.9 & 3.3 & 9.3 & 20.4 & 36.4 \\
  && 
  Spline & 0.0 & 0.1 & 1.1 & 2.1 & 5.4 & 8.6 & 15.0 \\ 
                  
    \midrule[1.25pt]
    
    $0.3$ &
    $0.15$ &
    
  Linear & 0.0 & 0.2 & 1.9 & 11.0 & 32.0 & 67.3 & 88.5 \\ 
  &&
  Spline & 0.0 & 0.5 & 1.5 & 6.3 & 13.5 & 26.0 & 39.8 \\ 
  & $0.23$ &  
  Linear & 0.0 & 0.4 & 1.7 & 10.4 & 29.0 & 58.0 & 80.6 \\ 
  &&
  Spline & 0.0 & 0.7 & 2.3 & 4.9 & 10.3 & 19.8 & 31.7 \\ 
  & $0.45$ &
  Linear & 0.0 & 0.4 & 1.1 & 3.4 & 11.0 & 23.3 & 37.2 \\ 
  &&
  Spline & 0.0 & 0.2 & 0.9 & 1.1 & 3.2 & 5.7 & 10.7 \\ 
                    
	\midrule[1.25pt]
	
	$0.45$ &
	$0.15$ &
	
  Linear & 0.0 & 0.1 & 1.1 & 9.1 & 28.9 & 65.2 & 89.2 \\ 
  &&
  Spline & 0.0 & 0.2 & 0.7 & 2.6 & 7.0 & 16.9 & 32.4 \\ 
  & $0.23$ &		
  Linear & 0.0 & 0.1 & 1.2 & 7.6 & 28.2 & 59.9 & 80.8 \\ 
  &&
  Spline & 0.0 & 0.2 & 0.7 & 2.5 & 5.3 & 13.4 & 24.1 \\ 
  & $0.45$ &
  Linear & 0.0 & 0.0 & 1.0 & 3.6 & 11.1 & 25.1 & 37.9 \\ 
  &&
  Spline & 0.0 & 0.0 & 0.4 & 0.7 & 2.1 & 3.2 & 8.8 \\ 
					                                          
  \bottomrule[1.25pt]
\end{tabular}
}
\end{table}

The probability that BFAST estimates a non existing abrupt change is nearly zero
even when 20\% of observations are missing. This feature changes as the amount
of missing values reaches 30\% (and onwards). In this situation, the false negative
probability is far smaller when spline-based fitting is used as an interpolation method
than when linear interpolation is employed as a gap filling procedure. These findings
are valid regardless bandwidth and amplitude values, see Table~\ref{tab_metodos_missVal_h_falseNegative}.

\subsection{Assessing BFAST's performance in estimating two abrupt changes}~\label{sec.sims.2CPS}

Here we are interested in assessing BFAST's ability to estimate two abrupt changes
as a function of the distance between them. We set $\sigma=0.02$, 
$h = 0.15, 0.23$, 
vary the percentage of missing data (from 40 to 60\%) and utilize a harmonic regression
model to simulate the seasonal component (amplitude $0.15$ and phase angle 0).
We focus on the behavior of BFAST when 40 to 60\% of observations are
missing in the time series based on the results of our previous simulation
and because this amount of missing information is relevant for our applications.

In this study the trend function is defined through Eq.~(1)
with $\alpha_1=0.7$, $\alpha_2=0.3$, $\alpha_3=\alpha_2$, $\beta_1=0$, 
$\beta_2=0.2/161$, $\beta_3=\beta_2$, $\tau_0=0$, $\tau_1=100$, 
$\tau_2=\tau_1+\ell$, and $\tau_3=322$.
Observe that the second abrupt change, $\tau_2$, is separated from the first one 
($\tau_1$) by $\ell$ time-points. In this study we use $\ell=10, 20, \ldots, 130, 140$. 

\begin{figure}[htb]
	\centering
		\includegraphics[width = 0.765\textwidth, height=0.6\textwidth]{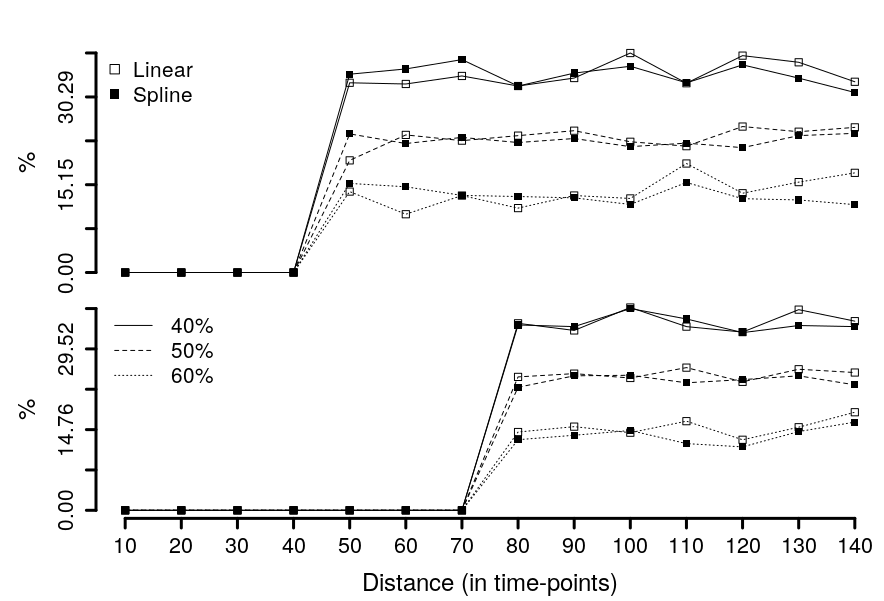}
	\caption{\footnotesize {Probability coverage. 
	Bandwidth: $h=0.15$ (top row), $h=0.23$ (bottom row).}}
	\label{FIG:3}
\end{figure}

We begin by studying BFAST's probability coverage of estimating \emph{correctly} 
the 2 true abrupt changes. That is, we divide the number of times in which BFAST 
estimates the true abrupt changes $\tau_1$ and $\tau_2$ by the number
of times in which 2 breakpoints are estimated. According to Figure~\ref{FIG:3},
the less percentage of missing data ($40\%$) the greater the BFAST's 
probability coverage, regardless of the interpolation method. Also from
that figure (top row, $h=0.15$) we infer that in order to obtain a non-zero
correct estimation probability is necessary that
the abrupt changes are separated by at least 50 time-points (approx.~2 years in
Landsat time scale);
this is in line with the fact that when $h=0.15$
the OLS-MOSUM statistic uses a bandwidth with roughly 48 time-points. Similarly,
(bottom row) when $h=0.23$ (and roughly 75 time-points are used in OLS-MOSUM's bandwidth)
BFAST begins to detect two abrupt changes as soon as they are separated by at least 80
time-points.

\begin{figure}[htb]
	\centering
		\includegraphics[width = 0.85\textwidth, height=0.8\textwidth]{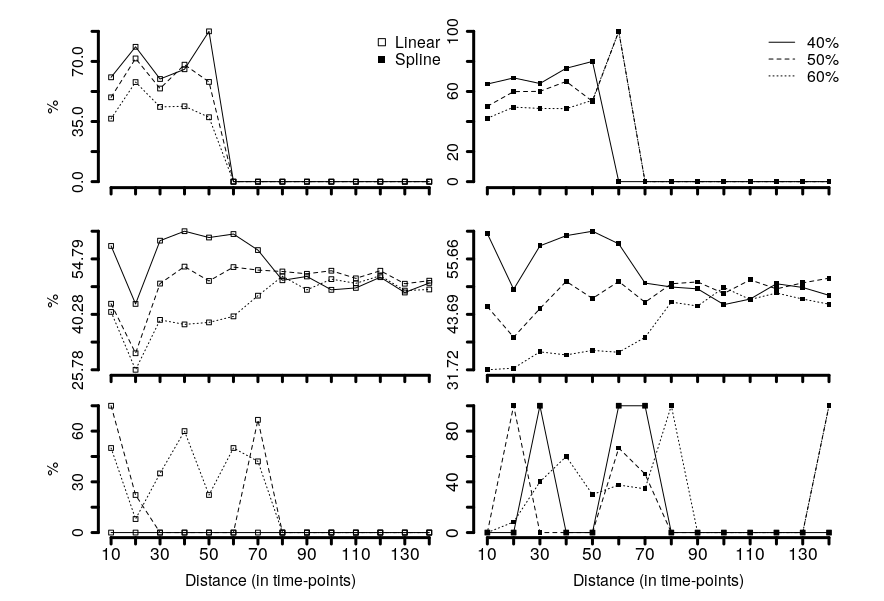}
	\caption{\footnotesize {Underestimation. 
	Bandwidth $h=0.23$. Interpolation methods: left column (linear), 
	right column (spline).}}
	\label{FIG:4}
\end{figure}

Next, we report on BFAST's underestimation, see Figure~\ref{FIG:4} for case $h=0.23$.
We consider underestimation, i.e.~either $\tau_1$ or $\tau_2$ are estimated
correctly,
when one (top row), two (middle row) or more than two (bottom row) breakpoints are 
detected. 
In the first case, and regardless of the interpolation method, when 
$40\%$ of data are missing, BFAST underestimates when the separation between
$\tau_1$ and $\tau_2$ is less than 60 time-points. This feature remains true
even when 50 and 60\% of observations are missing and when linear interpolation
is used as gap filling procedure.
In the second case (when two abrupt changes are detected), independently of the percentage 
of missing data and the interpolation method, BFAST's underestimation reaches
a stable level ($45\%$) once the separation is about 80 time-points.
Finally, when BFAST has detected more than two abrupt changes,
the linear interpolation method shows a less erratic behavior of the underestimation 
phenomenon throughout different amounts of missing data. Moreover, 
when 40\% of observations are missing and linear interpolation is used, BFAST reaches
zero underestimation probability coverage even when the separation between changes
is only 10 time-points.
Also, BFAST's underestimation will cease when the breakpoints are separated by 80 time-points.

\begin{figure}[hbt]
	\centering
		\includegraphics[width=0.85\textwidth]{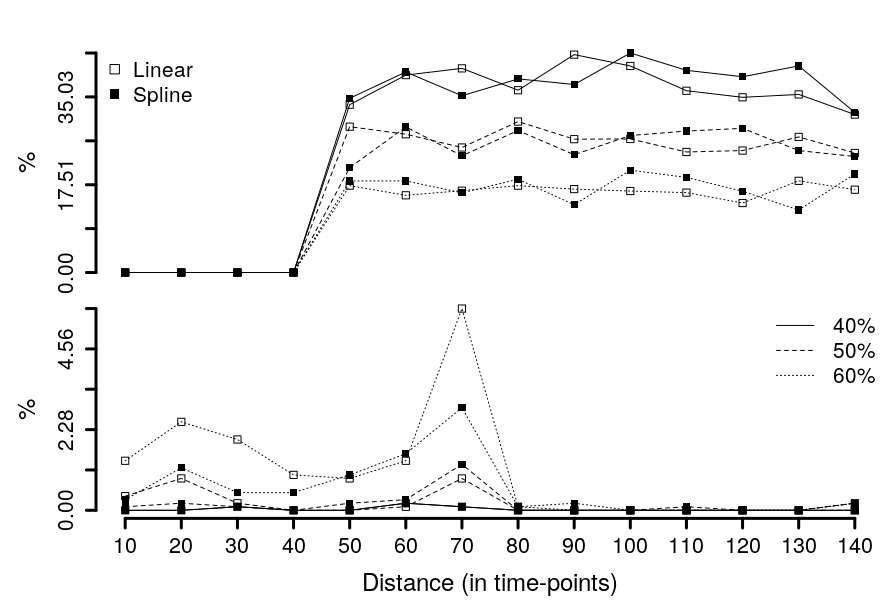} 
		\caption{\footnotesize {Overestimation. 
		Bandwidth: top row ($h=0.15$), bottom row ($h=0.23$).}}
		\label{FIG:5}
\end{figure}

From Figure~\ref{FIG:5} (bottom row) we conclude that
when the distance between two breakpoints
is at least 80 time-points and $h=0.23$,
BFAST does not show overestimation; this feature is independent
of the interpolation method and percentage of missing values.

\section{Conclusions}
The performace of BFAST to estimate abrupt changes deteriorates
as the amount of missing values in a time series increases.
All in all, BFAST's performance seems to be less affected by linear
than by spline interpolation. For one abrupt change: BFAST's correct estimation 
is greater with linear than with spline (even in the difficult case of lacking
$50\%$ observations); 
BFAST's accuracy and precision is better with linear interpolation
(when at least $30\%$ of values are missing); in contrast, 
the false negative probability is far smaller with spline interpolation.
For two abrupt changes linear interpolation shows a marginal better performance 
than spline interpolation.

As to for BFAST's bandwidth parameter, the use of $h=0.23$
precludes the detection of burned areas in 2005. Hence, in our
application we will also utilize the value $h=0.15$.

\ifCLASSOPTIONcaptionsoff
  \newpage
\fi

\end{document}